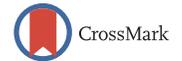

## Research Article
## Analysis of Data Mining Process for Improvement of Production Quality in Industrial Sector

Hamza Saad, Nagendra Nagarur and Abdulrahman Shamsan

Department of System Sciences and Industrial Engineering, Binghamton University, New York, USA

## Abstract

**Background and Objective:** Different industries go through high-precision and complex processes that need to analyze their data and discover defects before growing up. Big data may contain large variables with missed data that play a vital role to understand what affect the quality. So, specialists of the process might be struggling to defined what are the variables that have direct effect in the process. Aim of this study was to build integrated data analysis using data mining and quality tools to improve the quality of production and process. **Materials and Methods:** Data collected in different steps to reduce missed data. The specialists in the production process recommended to select the most important variables from big data and then predictor screening was used to confirm 16 of 71 variables. Seven important variables built the output variable that called textile quality score. After testing ten algorithms, boosted tree and random forest were evaluated to extract knowledge. In the voting process, three variables were confirmed to use as input factors in the design of experiments. The response of design was estimated by data mining and the results were confirmed by the quality specialists. Central composite (surface response) has been run 17 times to extract the main effects and interactions on the textile quality score. **Results:** Current study found that a machine productivity has negative effect on the quality, so this validated by the management. After applying changes, the efficiency of production has improved 21%. **Conclusion:** Results confirmed a big improvement in quality processes in industrial sector. The efficiency of production improved to 21%, weaving process improved to 23% and the overall process improved to 17.06%.

## INTRODUCTION

Data mining and Machine learning used extensively in industrial and business fields, helping to overcoming production issues and extract patterns from industrial data. The application strength of data mining in analyzing complex data has been proven in many studies[1,2]. The textile production is not new to the field of artificial intelligent, machine learning and data mining, however, it has different application areas in pharmaceutical, healthcare, financial, production, retail, administrative tasks and other business operations, such as the fortune 500 companies are tapped into the potential of data to gain deeper insight into their customer requirements. They even provide smart recommendations based on customers' past purchases. They work to gather data using machine learning models and use it to make important business decisions. For example, with the information gained from data, they can learn what products sell best and which ones need refining. Data mining can be also used by marketing teams in designing appealing and targeted promotions to attract more customers.

Companies report that they use a wide variety of data mining software. Oracle's Darwin, SAS's Enterprise Miner and IBM's Intelligent Miner are the dominant players, with SPSS's Clementine being used by a smaller number of companies. Different applications of information technology required to enhance productivity and get the best utilization along a production line. Successful organization is depending on the faster processing of raw data which strongly relies on advanced applications[3]. The use of applications for managing the quality and laboratory results of modern textile is insistent. Currently, most laboratories have some system to manage and log data. However, in the current competitive marketplace, profitability does not only depend on increasing sales but it also depends on improving quality and reducing costs. Introducing data mining techniques into the textile processes can give a substantial increase in the quality and productivity of the work. Data mining provides a set of techniques for studying the relationships and extract patterns from a huge dataset "this can be conducted automatically by identifying, validating and using data mining for prediction[4].

In textile manufacturing, even when a simple process or produce a standard product is considered, a big data is generated and stored. This data may include machine settings, raw materials and quality requirements. The data is complicated to handle if the relationship is required among process parameters, fiber properties and among yarn properties or yarn properties, fabric performance and machine settings[5]. Furthermore, there is difficult to define the goal when solving data of textile products especially if the product includes many specifications affect the quality. Although data mining has three analyzing techniques, regression, cluster and classification but not all techniques enable to deal with all textile data if these data collected randomly and directly from the production operations.

A lot of traditional statistical and mathematical models have been utilized in numerous textile studies to adjust and measure textile data[6,7]. However, these classical methods become incapable for discovering whole and complex relationships among the features or predicting unknown feature values for a new instance[8]. Because of this challenge, data mining that is conducted in a wide range of production and manufacturing areas has also been started used in textile engineering in recent years[5]. Data mining with its algorithms cannot offer an ultimate solution about the process but it can work with statistical analysis to provide a solution and apply statistical techniques to approve the possibility of this solution. In this study, algorithms of data mining work together to extract knowledge, then quality techniques will be applied to test these variables in the real process.

## MATERIALS AND METHODS

**Study area:** This study was conducted from April, 2019 to February, 2021 in Watson School, Binghamton University.

**Main idea:** Focusing on quality characteristics in the collection of variables may be difficult to search for the relationship between variables that may not be found in the same production line but they can be found in the whole production processes. Weaving process is affected by many factors that contribute to improving the quality. 71 variables were screened manually as a first step before filtering the variables using predictor screening.

The output variable called (Textile Quality Score) in the dataset was built carefully using the matrix of 7 quality characteristics, which is the integration of the seven important variables in the dataset and this will help to understand the important variables that affect the quality. Dealing with industrial data requires a greater challenge, especially with big data. Ten algorithms were tested in order to analyze the data and after comparing the results and determining the performance of the best algorithms, two algorithms were selected to analyze the data and extract the variables that affecting the textile quality score.

Data mining and quality specialists confirmed three variables that can be used in predicting the quality, we used these variables as factors in the design of experiments. Levels of factors picked from main dataset and the response variable has predicted using algorithms of data mining and get enough approving by quality specialists in the company to achieve the main contribution of this study by integrating data mining and quality techniques in real process to improve quality of product. The study development was outlined in Fig. 1.

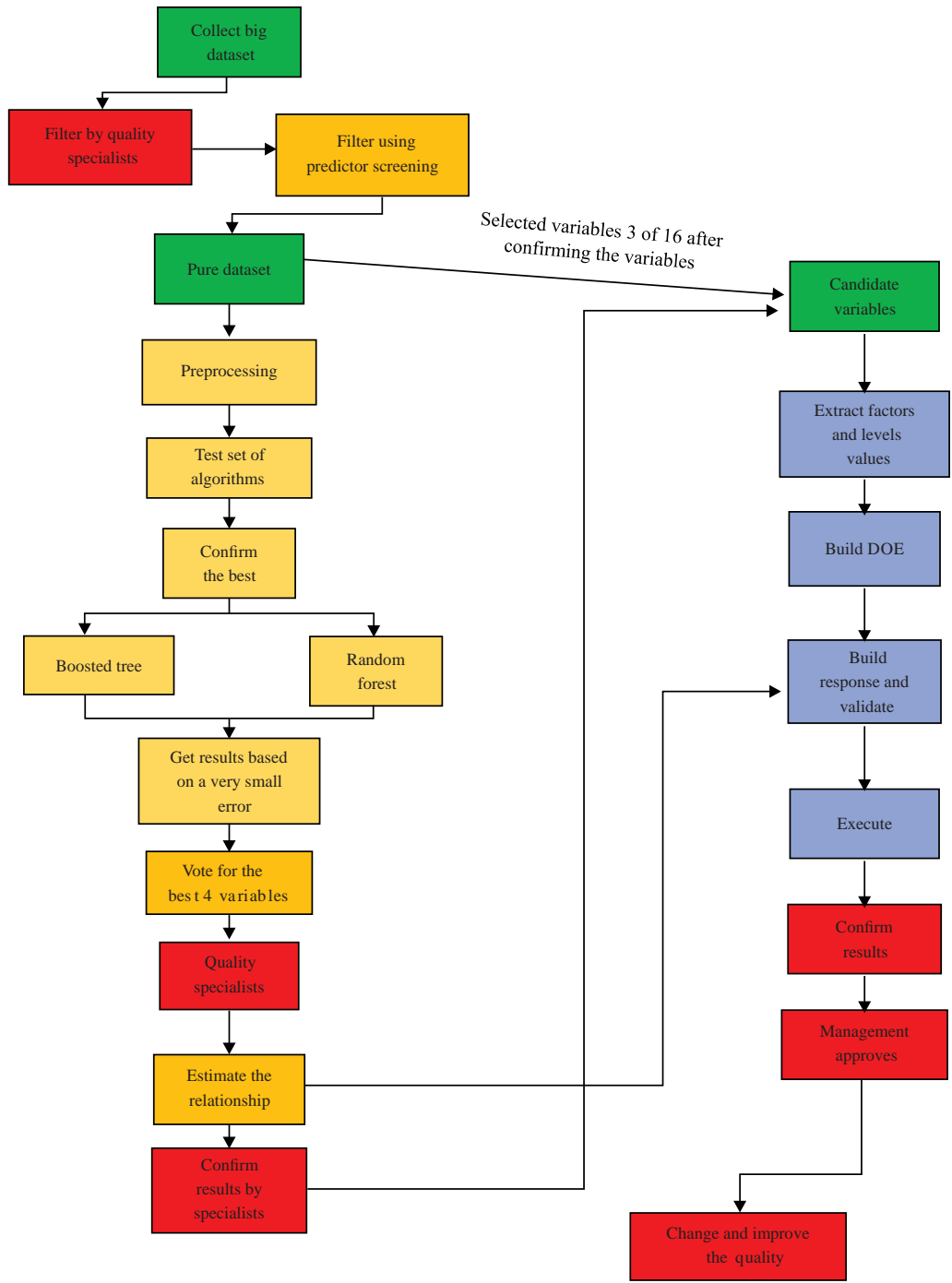

Fig. 1: Flow sheet diagram of study development

**Data gathering and preprocessing:** The operation process in the textile industry generates big data, there are important variables that control the quality of the product. The variables which pose direct and indirect relationship to the product quality were collected. The data was not ready for analysis, especially there is an inconsistency and a large data is missing (it is a form of data cleaning, where the purpose is to fill the variables that contain missed values with some intuitive data. In most of the cases, adding a reasonable estimate of a suitable data value is better than leaving it blank). To make valued intuition and organize the data, reference data from textile factories of Germany was used[9] which matched the quality specifications in order to fill important blanks and extract the related variables that are important in building the quality.

We stayed seven months working on collecting data and five months to validate study's results. Data was measured and picked in random processes (1-5) times to optimize and reduce high missed data and provide what can be used sufficiently in building the decision for improving product quality.

Using feature selection, 16 variables have approved to be handled into data mining algorithms. There are two types of variables, independent for one attribute and matrix by taking the average of variables between 2 and 5 (to be more accurate, some data was picked 5 times in order to extract variable with less blanks). These variables are explained based on the specialist of quality as following:

- **Humidity:** Controlled between 60 and 70% and the optimum humidity measured in the production processes is 65%, any change above this percentage will affect the process
- **Pigment fastness:** There are five tests for pigment fastness, if yarn or carpet failed a one test, then the product will miss some requirements of quality
- **Machine productivity:** There are two types of machine at the company: new and old. For new machines the productivity is between 87 and 93% and for old machines the productivity is between 45-71%. This variable is included worker efficiency, maintenance efficiency and shift operating loads
- **Cooling setting:** The optimum temperature measured to fit the operations is 25°C. If the temperature in the workplace is high, then the yarn will be easy to cut in the normal pulling
- **A number of stops per shift:** Stops come from low machine efficiency or miss control in humidity or temperature degree in the environment of the workplace
- **Designed color:** There are six colors designed for different textile products
- **Pile weight (gram/m²):** Textile yarn buys by the weight, therefore if pile weight is small, then it will be weak and easy to cut through the process. On another hand, if the pile weight is heavy, then it will cost much money

- **Pile height:** One of the quality characteristics of the final product specified by planned design
- **Ends per meter:** Designed manually and by a machine according to the required design
- **Tufts (m²):** Designed mechanically according to the detailed design
- **Total thickness (mm):** Thickness is required in planned design, if the thickness is less or high, then the quality will change
- **Oil contained:** This property is designed for the crude wool and the maximum oil allowed by the company is 0.19%
- **Surface height (mm):** This process designed manually and by a machine according to the required design
- **Carpet length:** The maximum length designed by machine is 25 m
- **Carpet width:** Managed by machine at 4 m for each produced carpet
- **Wool moisture contained:** A fundamental property because if a percentage exceeded 20%, then many problems might expose through the production processes
- **Dependent variable in the dataset is a textile quality score:** A number of specifications were carefully combined to give one percentage score (Pigment fatness, Strength, threat, Raw materials, Wool type, Textile weight and Pile weight)

**Predictor screening:** Predictor screening was used to get the reduction of the data set by removing redundant or irrelevant features (or dimensions). The goal of feature selection was to find a minimum set of attributes, such as the resulting probability distribution of the data output attributes, (or classes) is as close as possible to the original distribution obtained using all attributes. It increases the speed of the learning stage and facilitates the understanding of the pattern extracted[10].

The term curses of dimensionality[11,12] typically refers to the difficulties found in the fitting models, optimizing a function in many dimensions or estimating parameters. As the input data space dimensionality (the number of attribute predictors) increased, it becomes too difficult to get global optima for the parameter space (for cases instances, to fit the models). Virtually, the difficulties of certain models, for example, neural networks become impossible to manage when the number of input variables into the model exceeded a few hundred or even less. Therefore, it is necessary to choose and screen essential variables from a big set of predictor variables that are most likely used to predict the outputs (Textile Quality Score) of the interest.

The objective behind the Predictor Screening module is to select a set of predictor variables based on the dependent variable from a large list of candidates allowing to focus on a more professional set for further analysis. The algorithms of feature selection module were optimally handling categorical and continuous predictors, then estimate their predictive power that can improve accuracy and get a sophisticated output using influential predictor variables is so important.

Select features used to increase the accuracy by ignoring the low-efficiency variables which lead to reduce the accuracy of data analysis. 55 variables have been removed from dataset due to the low efficiency in predicting the selected output.

**Test for the best algorithms of data mining:** The effect of using each algorithm of data mining depends on the nature of the dataset, homogeneous or heterogeneous[13]. Many algorithms to solve this data were applied as shown in Table 1 and only two algorithms gave the flexibility and accuracy to deal with this data.

By testing a number of data mining algorithms, we verified only Random forests tree and Boosted tree which presented the highest flexibility and accuracy when applied to solve data.

**Random forests tree:** Random Forests consists of an ensemble or collection of simple tree predictors, each predictor can produce a response when presents with predictor values set. In classification problems, the response takes a form of membership class, where it classifies independent input values with one categorical dependent variable. On the other hand for regression problems, the tree response is the estimation of the dependent variable given predictors. The idea of the algorithm of Random Forest is developed by Breiman[14].

Random Forests tree includes arbitrary simple trees, which used to estimate the outcome. The basic improvements in the accuracy of classification or regression have been resulted from growing the ensemble trees by voting final results for the very popular classes. By increasing these ensembles trees, usually random vectors are generated and managed the growth of each tree in an ensemble technique. Bagging algorithm[14] where grows each tree in random selection (without replacement) found from the example in the training set.

Table 1: Results and some details from data mining algorithms

| Algorithm | MSE | RMSE | MAE | Note | Why? |
|---|---|---|---|---|---|
| Neural network | 0.025 | 0.009 | 0.180 | X | X |
| Tree C4.5 | 0.020 | 0.143 | 0.110 | X | X |
| KNN | 0.014 | 0.117 | 0.096 | X | X |
| Naïve Bayes | X | X | X | It does not fit to solve this type of data | Naive Bayes is used for strings and numbers (categorically), it can be used for classification so that it can be whether 1 or 0 nothing in between like 0.5 (regression) |
| Chi-squared automatic interaction detector (CHAID) | 0.0103 | 0.001 | X | It does not estimate important variables correctly | It does not fit entirely to data of the study |
| Classification and regression tree (CART) | X | X | X | The algorithm cannot use cross-validation or any data split to solve this type of data | It is ranked the variables based on which one is more significant, but its result was not accepted because the variables in real production cannot make the improvements or changes |
| Association rules | X | X | X | It does not fit to solve this type of data | It applied for classification. There are no categorical variables in this data |
| K-means | X | X | X | It does not fit for the regression problem | Data needs to solve in regression, not in clustering, so no categorical output in this data |
| Boosted tree | 0.009 | 0.0016 | X | It is very suitable | It can deal with data in a smoothly way |
| Random forests | 0.012 | 0.110 | 0.091 | It is appropriate for this data | It extracted the important variables with low error |

MSE: Mean square error, RMSE: Root mean square error and MAE = Mean absolute error

14

Another example, random split selections[15] where at each node split selected at random from among the K best splits. Breiman[14] generated a new training set by randomizing the output in the base of the training set. Another method selected is the training set from the random weights set in the example of the training set.[16] wrote few papers on "Random Subspaces" approach which made a random selection of features subset to use for growing each tree.

In the important study on written character recognition,[17] defined a large number of geometric features and they search over random selection for these to get the best split at each node.

Element used in most procedures is for the $K^{th}$ trees. The random vector $Q_k$ is obtained independently from the previous random vectors $Q_1,...,Q_{k-1}$ have the same distribution as a condition. The tree is grown using the $Q_k$ and training set resulted in the classifier $h(x, Q_k)$ where x is the input vector. For example, bagging a random vector Q is generated as counts in N boxes resulted from N data goes to random boxes, where N represents examples number in the training set. In the random split, the selection of some independent of the random integers between K and 1. After generating a number of trees, they are voting for the most popular class, these processes called a random forest. By applying this algorithm, we got small errors shown in Table 2.

**Boosted tree:** he algorithm of boosting trees was developed from the applications of boosting techniques for regression trees. The main notion is for computing the sequence of (very) simple trees, where each successive tree is established for the prediction residual of the preceding tree[18,19]. As in the General Classification and Regression Trees, this technique built on binary trees as an example, the partition of the data[20] into two samples at each split node. Now suppose that to limit the complexity of the trees to only three nodes: Two child nodes and a root node, i.e., a single split. Therefore, at each step of a boosting (boosting trees algorithm), a simple or (best) partitioning of the data is defined and the deviations of observed values from the respective means (residuals for each partition) are calculated. In the next three, nodes will be fitted to those residuals. To find another partition which algorithm gives further reduce a residual (variance) error for the data, given the trees preceding sequence.

Such "additive weighted expansions" of the trees can finally generate the excellent fit of the observed values to the predicted values, even if the precise nature of the relationships between the dependent variable and the predictor variables of the unusual case is a very complex or (non-linear)[21]. Hence, the method technique of the gradient boosting-fitting the weighted additive expansion of the simple trees represented a very powerful and general machine learning algorithm.

The common computational technique of stochastic gradient boosting is also called by the names of MART (TM Jerill, Inc.) and TreeNet (TM Salford Systems, Inc). In the past few years, this approach has presented as one of the most powerful techniques for data mining prediction. Some applications of these algorithms allowed them to be used for classification as well as regression problems, with categorical and continuous predictors. The standard error and risk estimate were shown in Table 3.

Each algorithm has ranked variables based on the impact on the dependent variable (textile quality score). According to the results of algorithms, four best variables were selected for the next step. Table 4 presented the variables selected from data mining.

Table 2: Error estimation using random forests

| Random forests | Risk estimate | Standard error |
|---|---|---|
| Train | 0.008030 | 0.000965 |
| Test | 0.013378 | 0.002315 |

Risk estimates are forecasts of the probability and impact of risks. They are used to manage risks. Standard error is the standard deviation of its sampling distribution or an estimate of that standard deviation

Table 3: Error estimation using boosted tree

| Boosted tree | Risk estimate | Standard error |
|---|---|---|
| Train | 0.009055 | 0.001053 |
| Test | 0.009025 | 0.001676 |

Table 4: Voting process

| Predicted Inputs for the best four rank variables (random forests) | Predicted inputs for best four rank variables (boosted tree) | Voted prediction for the best voted variables |
|---|---|---|
| Machine productivity (%) | Tufts ($m^2$) | Tufts ($m^2$) |
| Tufts ($m^2$) | Pigment fastness | Machine productivity (%) |
| Pile height | Machine productivity (%) | Pile weight (g $m^{-2}$) |
| Pile weight (g $m^{-2}$) | Pile weight (g $m^{-2}$) | Pigment fastness |

**Vote for the influential variables based on the bagging technique:** In variables ranking, there is a big gap between some variables of the dataset and that explains this each algorithm splits data in different processing: Random Forest used a Bootstrap (Sample with the replacement) and Boosted used weighting to the weak classifiers. Three input variables (machine productivity, pile weight and Tufts ($m^2$)) have chosen directly due to high ranking in both algorithms. Pigment Fastness variable in real manufacturing is more important than the variable of Tufts ($m^2$) according to the production and quality management department in the Company. Variables were confirmed by a real decision from the Company after voting selected variables as what presented in Table 4.

Machine productivity and pile weight variables got the highest ranking and appeared in the final result of each algorithm. However, Pigment fatness was used instead of a variable of Tufts ($m^2$) as the third variable in the design of experiments. The reasons behind that were, (1. pigment fastness is more important than Tufts ($m^2$) for quality improvement and (2. Tufts ($m^2$) is modified by machine based on the operation plan and if the operation process get a perfect check, then Tufts ($m^2$) will be in standard requirements. Final variables selection was evaluated by the management of the Company, which approved only three variables to improve the final product. By applying data mining, we estimated this equation (Textile Quality Score = 0.896502+(0.067231*Pigment Fastness)-(0.1482945*Machine Productivity)+(0.000005*Pile Weight)) to help quality specialists in defining the response of textile quality score based on the defined factors.

**Conduct the design of experiments based on the voted variables:** In the design of experiments, three factors have been selected from data mining after they got enough testing by professional workers. Each factor has three levels. According to each factor, the highest, medium and lowest value have been picked from the original dataset which were presented in Table 5.

By applying these levels, design is conducted and validated for 17 runs and solved using the response surface experiments. The response of (Textile Quality Score) is not entirely accurate because of engineers unable to give exact responses due to difficult to run real experiments. Data minig used to solve the problem and estimate the responses by focusing on selected variables. The result of prediction was submitted to professional workers in the company to build an exact response for each run based on the real process. Table 6 presented estimate and validate the response.

Table 5: Levels of each factor

| Levels/factors | Pile weight | Machine productivity | Pigment fasting |
|---|---|---|---|
| High level | 2729.0 | 0.93 | 1.0 |
| Medium level | 2114.5 | 0.69 | 0.875 |
| Low level | 1500.0 | 0.45 | 0.75 |

Table 6: Estimated and Validated Values

| Estimated by data mining | Validated by professional workers |
|---|---|
| 0.9045 | 0.905 |
| 0.887693 | 0.89 |
| 0.816511 | 0.816 |
| 0.863578 | 0.864 |
| 0.833319 | 0.833 |
| 0.860506 | 0.861 |
| 0.910645 | 0.911 |
| 0.839464 | 0.84 |
| 0.827988 | 0.83 |
| 0.866651 | 0.87 |
| 0.863578 | 0.864 |
| 0.871982 | 0.872 |
| 0.822656 | 0.823 |
| 0.893838 | 0.894 |
| 0.863578 | 0.864 |
| 0.899169 | 0.9 |
| 0.855175 | 0.8552 |

Response Surface design in Table 7 was constructed based on three blocks, three levels, three factors and 17 runs.

**Statistical analysis:** To extract more knowledge from our dataset and build the improvement in the process we applied design of experiments (Response Surface Design). It is a collection of statistical and mathematical techniques for building an empirical model. The objective of DoE is to optimize responses (output variables) which is affected by different independent variables (input factors)[22]. The experiments are a series of tests, named runs, in which change is conducted in the input factors to recognize the main reasons for changes in the output responses. One of DoE methods is Response Surface Methodology (RSM).

By applying Non-factorial (Response surface) design in order to get the most important variables effect in the textile quality score, based on the analysis we can rank the highest significant in the design as in Table 8.

## RESULTS

**Effect estimate:** All factors have a significant linear effect on the output but the rank of the effect is changing based on the impact of each factor. The factor of machine productivity has a negative and robust effect and the factors of pigment fastness and pile weight have a positive effect but the impact of pile weight is the lowest among all factors. There is a weak quadratic effect but there is a clear linear effect as shown in Table 9.

Table 7: Response surface design with one response

| Standard run | Blocks | Pigment fastness | Machine productivity | Pile weight | Textile quality score |
|---|---|---|---|---|---|
| 8 | 2 | 1.000000 | 0.450000 | 1500.000 | 0.905 |
| 1 | 1 | 0.750000 | 0.450000 | 1500.000 | 0.89 |
| 7 | 2 | 0.750000 | 0.930000 | 1500.000 | 0.816 |
| 5 (C) | 1 | 0.875000 | 0.690000 | 2114.500 | 0.864 |
| 4 | 1 | 1.000000 | 0.930000 | 1500.000 | 0.833 |
| 15 | 3 | 0.875000 | 0.690000 | 1500.000 | 0.861 |
| 3 | 1 | 1.000000 | 0.450000 | 2729.000 | 0.911 |
| 9 | 2 | 1.000000 | 0.930000 | 2729.000 | 0.84 |
| 14 | 3 | 0.875000 | 0.930000 | 2114.500 | 0.83 |
| 16 | 3 | 0.875000 | 0.690000 | 2729.000 | 0.87 |
| 10 (C) | 2 | 0.875000 | 0.690000 | 2114.500 | 0.864 |
| 12 | 3 | 1.000000 | 0.690000 | 2114.500 | 0.872 |
| 2 | 1 | 0.750000 | 0.930000 | 2729.000 | 0.823 |
| 6 | 2 | 0.750000 | 0.450000 | 2729.000 | 0.894 |
| 17 (C) | 3 | 0.875000 | 0.690000 | 2114.500 | 0.864 |
| 13 | 3 | 0.875000 | 0.450000 | 2114.500 | 0.9 |
| 11 | 3 | 0.750000 | 0.690000 | 2114.500 | 0.8552 |

Table 8: Rank the most important factors

| Factor | Rank | Type of relationship |
|---|---|---|
| Machine productivity | 1 | Strong negative |
| Pigment fastness | 2 | Weak positive |
| Pile weight | 3 | Weak positive |

Table 9: Effects estimation

| Factor | Effect | Standard error | t (5) | P |
|---|---|---|---|---|
| Mean/interc. | 0.864126 | 0.000580 | 1489.964 | 0.000000 |
| Blocks (1) | -0.000840 | 0.000814 | -1.033 | 0.348921 |
| Blocks (2) | 0.000881 | 0.000902 | 0.977 | 0.373449 |
| (1) Pigment fastness (L) | 0.016560 | 0.000677 | 24.454 | 0.000002 |
| Pigment fastness (Q) | -0.002121 | 0.001352 | -1.568 | 0.177558 |
| (2) Machine productivity (L) | -0.071600 | 0.000677 | -105.730 | 0.000000 |
| Machine productivity (Q) | 0.000679 | 0.001352 | 0.502 | 0.637131 |
| (3) Pile weight (L) | 0.006600 | 0.000677 | 9.746 | 0.000193 |
| Pile weight (Q) | 0.001679 | 0.001352 | 1.241 | 0.269599 |
| 1L by 2L | 0.000500 | 0.000757 | 0.660 | 0.538208 |
| 1L by 3L | 0.000500 | 0.000757 | 0.660 | 0.538208 |
| 2L by 3L | 0.001000 | 0.000757 | 1.321 | 0.243793 |

**ANOVA analysis:** Based on the p-Value from Table 10, all factors are significant. So, the machine productivity factor has the most significant impact on the design based on p-value that equals 0.

According to Fig. 2, Machine Productivity has the most significant effect on the design. All factors gave a significant effect on the output but they have different scales. For example, the machine productivity factor has a negative effect, the pigment fastness factor and pile weight factor have a positive impact.

The three variables make a change in the production process, they contributed to improving the efficiency by 17.06% after five months based on the qualified specialists in the company who checked and evaluated the production reports.

Machine productivity is related negatively to the textile quality score and this was the problem. We found in the company that old machines with low efficiency. For example, in the summer, there are more cuts in the threads as a result of the weaving process of pulling the threads in the machine and the workers cannot stop the production process because it will cost more money but they tied the threads during operation, which affect the quality. To avoid this problem, we need to optimize the machine operation and workplace temperature using air-conditioners to reduce the cutting in the threads, which increases production costs.

Many stops happen to machines due to static electricity that may lead to death, the static electricity is generated from

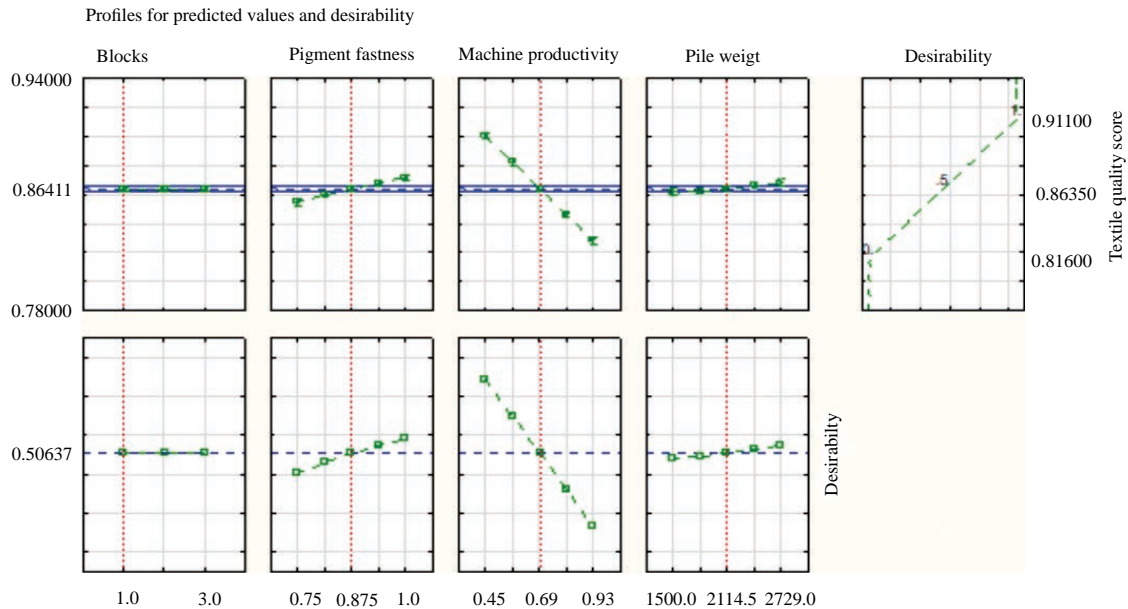

Fig. 2: Predicted values vs. desirability

Table 10: ANOVA analysis

| Factor | SS | df | MS | F | P |
| --- | --- | --- | --- | --- | --- |
| Blocks | 0.000001 | 2 | 0.000001 | 0.65 | 0.560391 |
| (1) Pigment fastness (L) | 0.000686 | 1 | 0.000686 | 597.99 | 0.000002 |
| Pigment fastness (Q) | 0.000003 | 1 | 0.000003 | 2.46 | 0.177558 |
| (2) Machine productivity (L) | 0.012816 | 1 | 0.012816 | 11178.90 | 0.000000 |
| Machine productivity (Q) | 0.000000 | 1 | 0.000000 | 0.25 | 0.637131 |
| (3) Pile weight (L) | 0.000109 | 1 | 0.000109 | 94.99 | 0.000193 |
| Pile weight (Q) | 0.000002 | 1 | 0.000002 | 1.54 | 0.269599 |
| 1L by 2L | 0.000000 | 1 | 0.000000 | 0.44 | 0.538208 |
| 1L by 3L | 0.000001 | 1 | 0.000001 | 0.44 | 0.538208 |
| 2L by 3L | 0.000002 | 1 | 0.000002 | 1.74 | 0.243793 |
| Error | 0.000006 | 5 | 0.000001 | 0.65 | 0.560391 |
| Total SS | 0.013625 | 16 | | | |

SS: Sum of square, DF: Freedom degree, MS: Mean square and F: F score measure weaving process and its strength increases when touch the metal body of the machine, which increases its electricity power and influence on the workers who are near it and as result this will lead to an emergency stop and a negative effect on the production process and the quality.

Also, workers operate the machines in the wrong way by carrying the machines more than their capacity to achieve required demands. The workers also do not allocate work hours and stoppages for machines in the right way.

We took five months to apply some actions in order to change management and improve the efficiency of production. These actions were summarized in Table 11.

## DISCUSSION

The design of experiment helps to extract the main variables that have a strong relationship with the textile quality score[22]. By using machine productivity, pile weight and pigment fastness as factors in the DOE, we found that all factors are significant with the different impact levels, so machine productivity was related to textile quality score with the highly significant. We sent these results to the Company and they validated the factor of machine productivity that needs work hard in order to make the improvement in the production and process. The results from the design of experiments have been extracted from main effects analysis for each factor, ANOVA analysis[23] and predicted variables VS desirability.

We built a reliable system to analyze data in the production process by considering all the variables that are related to the production process[9]. Boosted tree[18] and Random Forests tree in data mining[1] cannot give a full improvement to the process without getting support from the design of experiments or other quality tools. Data mining was filtered[10] the most important variables that strongly related to the textile quality score[5]. The best tool that can be used to extract the relationships, interactions and the main effect is the design of experiments. So, this tool is insufficient to deal with big data or huge numbers of factors, that is why we thought to integrated data mining with the design of experiments through multiple linear regression.

The limitation of data analysis used in this strategy is that the strategy cannot be fitted with categorical variables, inconsistent datasets or nonlinear relationships with high standard error. They can be used independently without integrating with quality tools.

## CONCLUSION

Improving product quality begins with an integrated quality system that should design and prepare by skilled employees in the company, otherwise, problems could appear in the final product especially in the textile manufacturing. Current study understands the system in order to get the problem points in the whole production process to apply the corrective actions in the process. There is an emphasis on a limited number of machines in the production line. That is, the higher operating load on these machines. Changes were applied by re-distributing machines, distributing operating loads and adding new machines to the production line to increase the efficiency.

Each production unit includes skilled workers who got enough training to apply production and quality plan assigned to supervise and ensure product quality, especially in the final stages. After five months, we analyzed the production reports by quality specialists, to find out that production efficiency has improved 21%, productivity improved 23% after fixing bottlenecks in weaving unit and quality has improved by reducing defects because of machines.

## SIGNIFICANCE STATEMENTS

Data analysis in industrial fields is so important to maintain the continuity of industrial operations. The manufacturing data in the raw state is huge, heterogeneous or non-linear, which increases the complexity if we used the traditional quality tools. However, quality techniques like the design of experiments are more effective in the case of small data. Therefore, we applied the full strategy of data mining in order to extract the most related data to facilitate the application of the design of experiments to find the appropriate improvement in the process.

Table 11: Recommendations for corrective action

| Process | Action (what we exactly did) | Action | Result (after 5 months) (02/2020 to 07/2020) |
|---|---|---|---|
| Worker efficiency | Provide training policy designed to the process and hire skilled workers | Done | The efficiency has improved |
| Efficiency of production | Reduce distractions, practice positive reinforcement and provide right tools and equipment for right workers | Done | Efficiency of production has improved 21% |
| Machines | Minimizing machine malfunction, optimize operating loads and encourage only skilled workers to operate machines | Done | Machines became more productivity |
| Quality defects | Set the tolerances for quality defects in a quality manual, encourage the workers to meet the quality requirements and build quality in the first step | Done | Defects of quality were reduced |
| Product | Produce products according to market demands, focus on one type of product, design and develop product that is more attracted to the customers | Done | Product became more distributed in the local market, but for one type of product, so company has five types of products need to distribute later |
| Bottlenecks | Define bottleneck points, increase the flexibility, define value-add activities, solve problems of production before starting and make sure there are no stops in the operations during the shift | Done | By fixing weaving bottlenecks, we got improve in the productivity to 23% |
| Efficiency of processes | Measure the baseline process, define the resources to improve the efficiency, provide a perfect work time, optimize the loads on machines and select skilled workers to lead the process | Done | The process has improved because of the skilled workers to 17.06% |